\documentclass[preprint,pre,oneside,floatfix]{revtex4-1}

\usepackage{amsmath}
\usepackage{amssymb}
\usepackage{graphicx}
\usepackage{dcolumn}
\usepackage{bm}
\usepackage{color}
\usepackage{wasysym}

\begin{document}


\title{Spontaneous symmetry breaking of charge-regulated surfaces}

\author{Arghya Majee$^{1,2}$}
\email{majee@is.mpg.de}
\author{Markus Bier$^{1,2}$}
\email{bier@is.mpg.de}
\author{Rudolf Podgornik$^{3,4}$}
\email{rudolf.podgornik@ijs.si}
\affiliation
{
$^{1}$Max Planck Institute for Intelligent Systems, 
      Heisenbergstr.\ 3,
      70569 Stuttgart,
      Germany\\ 
$^{2}$Institute for Theoretical Physics IV,
      University of Stuttgart,
      Pfaffenwaldring 57,
      70569 Stuttgart,
      Germany\\
$^{3}$Department of Theoretical Physics, 
      J.\ Stefan Institute,
      Jamova c.\ 39,
      1000 Ljubljana, 
      Slovenia\\
$^{4}$Department of Physics, 
      Faculty of Mathematics and Physics, 
      University of Ljubljana, 
      Jadranska 19,
      1000 Ljubljana, 
      Slovenia
}

\begin{abstract}
The interaction between two \textit{chemically identical} charge-regulated surfaces is studied 
using the classical density functional theory. In contrast to common expectations and assumptions, 
under certain realistic conditions we find a spontaneous emergence of disparate charge densities 
on the two surfaces. The surface charge densities can differ not only in their magnitude, but 
quite unexpectedly, even in their sign, implying that the electrostatic interaction between the 
two chemically identical surfaces can be \textit{attractive instead of repulsive}. Moreover, 
an initial symmetry with equal charge densities on both surfaces can also be broken spontaneously 
upon decreasing the separation between the two surfaces. The origin of this phenomenon is a 
competition between the adsorption of ions from the solution to the surface and the interaction 
between the adsorbed ions already on the surface.These findings are fundamental for the 
understanding of the forces between colloidal objects and, in particular, they are bound to 
strongly influence the present picture of protein interaction.
\end{abstract}

\maketitle


\section{Introduction} 
Within the mean-field Poisson-Boltzmann (PB) paradigm of the electrostatic 
interaction between two charged surfaces immersed in an ionic solution, one usually assumes a constant surface charge density
or a constant surface potential boundary conditions \cite{Saf18}. 
Although this simplifies the problem, most common naturally occurring nanoparticle and macromolecular surfaces of 
interest, e.g., hard colloidal particles, soft biological molecules including proteins, membranes, 
and lipid vesicles rarely satisfy either of them \cite{Pop10, Lun13}. 
They respond to their environment, especially to the presence of each other, in such a way that both the
charge density and the surface potential vary and adjust themselves 
to the separation between the surfaces as well as to the bathing solution environment. This conceptual framework 
is formally referred to as the \textit{charge regulation} mechanism and can be formalized either by invoking the 
chemical dissociation equilibrium of surface binding sites with the corresponding law of mass action, 
an approach pursued in the seminal work of Ninham and Parsegian \cite{Nin71}, or equivalently by adding a 
model surface free energy to the PB bulk free energy that \textit{via} minimization then leads to the same basic 
self-consistent boundary conditions for surface dissociation equilibrium but without an explicit connection 
with the law of mass action \cite{diamant, Hen04, Olvera2, maggs}. The latter approach is to be preferred when the 
surface dissociation processes are more complicated, as discussed below, and can not be captured by a simple 
law of mass action.

Studies of the interaction between two charge-regulated surfaces have been performed for
chemically identical surfaces with equal adsorption and desorption properties 
\cite{Beh99_JPCB, Bie04, Bor08, Tre16} as well as for chemically non-identical surfaces 
\cite{Cha76, Mcc95, Beh99_PRE, Cha06, Pop10}. In all of the former cases, a certain basic symmetry of the 
problem was assumed \textit{a priori} \cite{Sad99, Neu99, Tri99} and the surface charge densities 
have been without exception constrained to be equal on both surfaces. However, the underlying physical 
reasoning for such an assumption is not generally applicable and is not based upon 
the detailed chemical nature of the surfaces bearing charge. 
The fact that the two interacting surfaces are chemically identical and, therefore, interact in the same 
way with the adjacent bathing solution, is \textit{not sufficient} to infer the fundamental charge symmetry 
and to invoke equal surface charge densities in the application of the PB formalism. 
In fact, the charge distribution of the system should not be assumed \textit{a priori}, but should follow from 
the minimization of the relevant total thermodynamic potential, yielding the equilibrium state in terms of the 
equilibrium electrostatic potential distribution between the surfaces as well as the equilibrium charge 
densities on the surfaces, without any additional assumptions.
Whether this minimum implies an equal or unequal surface charge densities may and, as will be 
shown below, does depend on the parameters of the system under consideration.

Below we show that, depending on these parameters, a confined electrolyte in thermodynamic equilibrium 
with two chemically identical charge-regulated surfaces that can adsorb/desorb solution ions, can 
indeed adopt \textit{unequal surface charge densities}, even at separations that are much larger than the 
Debye length. This happens due to an interplay of the adsorption of ions from the solution \textit{to the surface} 
and the interaction between the already adsorbed ions \textit{at the surface}. The model surface free energy 
is then related to the lattice fluid model and is composed of the surface entropy of mixing, the electrostatic 
energy of the adsorbed charges, the non-electrostatic energy penalty of adsorption and the change in the 
non-electrostatic interactions between the ions upon adsorption. The latter are assumed to be short-ranged, 
typically of the van der Walls  type, hydrogen-bonding and/or of quantum-chemical origin, 
that allow for a nearest-neighbor-like description. Since in general the surface charge densities can differ 
in magnitude as well as in sign, an initial symmetry with equal charge densities on both surfaces can be 
spontaneously broken, and the surfaces can acquire different charge densities as they approach each other. 
At short separations, this implies surface charge densities differing even in their sign and consequently leading 
to an overall attractive interaction between the surfaces. An analytical treatment of the simpler system with 
only a single surface in contact with an electrolyte indicates that these findings are inherently related solely 
to the electrostatic interaction between the surfaces.

\begin{figure}[!t]
\centering
  \includegraphics[width=7cm]{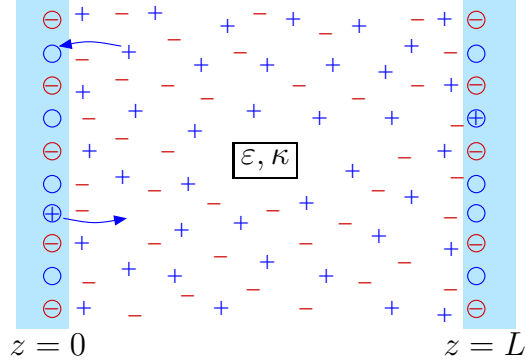}
  \caption{Schematic illustration of two chemically identical surfaces separated by a distance $L$ and 
           interacting electrostatically across an electrolyte solution with permittivity $\varepsilon$ and
           inverse Debye length $\kappa$. Both surfaces contain fixed negative surface charges \textcolor{red}{$\ominus$} 
           and sites \textcolor{blue}{$\fullmoon$} where cations (\textcolor{blue}{\footnotesize{+}}) from the 
           solution can attach or from which adsorbed cations \textcolor{blue}{$\oplus$} can detach.}
  \label{Fig1}
\end{figure}

\section{Model} 
There are several models present in the literature, based on the surface free energy implementation of 
the charge regulation process, describing, e.g., of mineral surfaces \cite{Hie_89_1, Hie_89_2, Kos05} or 
lipid membranes \cite{Har06, Mar16}, and here we follow the latter. We consider two charge-regulated, 
chemically identical planar surfaces situated perpendicular to a $z$-axis at positions $z=0$ and $L$ with 
an electrolyte solution in between (see Fig.~\ref{Fig1}). Each surface contains a fixed number of negative 
charges per surface area, $N$, and a number of neutral sites per surface area, $\Theta N$, where adsorption 
and desorption of cations can take place, leading to charge regulation of the surfaces. The charge density 
on a surface is then given by $\sigma=-Ne+Ne\eta\Theta$ with $e>0$ being the elementary charge and $\eta$ 
denoting the fraction of occupied sites on a surface. Since by construction $\eta\in[0,1]$, for $\Theta=2$, 
i.e., when there are twice as many sites present compared to fixed negative charges \cite{Nat15}, the charge 
density varies within a symmetric interval $\sigma\in[-Ne,Ne]$. The area per site is $a^2=1/(\Theta N)$, and 
we define the dimensionless charge density as 
\begin{align}
 \sigma^{*}=\frac{a^2\sigma}{e}=\eta-\frac{1}{\Theta}.
 \label{eq1}
\end{align}
The parameter $\eta$ and therefore $\sigma^*$ are assumed to be uniform over the surface. 
The electrolyte is considered to be a structureless, linear dielectric medium with permittivity 
$\varepsilon=\varepsilon_r\varepsilon_0$, where $\varepsilon_0$ is the permittivity of the vacuum and
$\varepsilon_r$ is the relative permittivity. The solute is a monovalent salt of bulk ionic strength $I$ 
and the corresponding Debye screening length is given by $\kappa^{-1}=\sqrt{\varepsilon/(2\beta e^2I)}$ with 
the inverse thermal energy $\beta=1/(k_BT)$.

\section{Density functional theory} 
Considering the bulk of the electrolyte as a reservoir for the ions, treating them as point-like 
particles, and ignoring ion-ion correlations within a mean-field formalism, the grand potential 
$\Omega\left[\sigma^*,n_{\pm}\right]$ can be written in terms of the number density profiles of ions 
$n_\pm\left(\mathbf{r}\right)$ and surface charge density $\sigma^*(\mathbf{r})$. The Euler-Lagrange 
equation minimizing this grand potential with respect to $n_{\pm}$ leads to the PB equation 
$\nabla^2\psi=\kappa^2\sinh\psi$ subjected to Neumann boundary conditions at the surfaces set by 
$\sigma^*$. Here $\psi(\mathbf{r})$ is the dimensionless electrostatic potential expressed in units 
of $\beta e$. Hence the equilibrium ion number density profiles $n_{\pm}\left[\sigma^*\right]$ and the 
equilibrium electrostatic potential $\psi\left[\sigma^*\right]$ are functionals of $\sigma^*$. 
Inserting $n_{\pm}\left[\sigma^*\right]$ in the expression for $\Omega\left[\sigma^*,n_{\pm}\right]$ 
one obtains the total grand potential functional in terms of the surface charge density profile $\sigma^*$. 
In the present work the surface charge density profile $\sigma^{*}$ is  assumed to be laterally 
uniform on each surface, i.e., $\sigma^{*}\left(\mathbf{r}\right)$ and consequently, $\eta\left(\mathbf{r}\right)$ 
may depend at most on 
$z = 0, L$. As a result, the electrostatic potential $\psi$ also depends 
on the $z$-coordinate only. With these, $\widetilde\Omega\left[\sigma^*\right]$, which is the grand potential 
functional $\Omega\left[\sigma^*\right]$ per unit surface area, corresponding to our system is given by
\begin{align}
 \beta\widetilde\Omega\left[\sigma^*\right] 
     =&\frac{-\varepsilon}{\beta e^2}\int\limits_{0}^{L}dz\Bigg[\kappa^2\cosh\left(\psi\left(z,\left[\sigma^*\right]\right)\right)
     +\frac{1}{2}\left(\psi'\left(z,\left[\sigma^*\right]\right)\right)^2\Bigg]\notag\\
   & +\frac{1}{a^2}
   \sum\limits_{z=0,L}\Bigg[\sigma^*\left(z\right)\psi\left(z,\left[\sigma^*\right]\right)\notag\\
   & -\alpha\eta\left(z\right)
     -\frac{\chi}{2}\eta\left(z\right)^2\notag\\
   & +\eta\left(z\right)\ln\eta\left(z\right)
     +\left(1-\eta\left(z\right)\right)\ln\left(1-\eta\left(z\right)\right)\Bigg],
\label{eq2}
\end{align}
where $\psi'\equiv\partial_z\psi$ and $\eta\left(z\right)=\sigma^*\left(z\right)+1/\Theta$ according to Eq.~(\ref{eq1}). 
The first line of Eq.~(\ref{eq2}) represents the volume electrostatic contribution to the grand potential and is 
identical to the standard PB form \cite{Saf18}. The second line represents the standard surface electrostatic energy 
of the adsorbed charges. The third line describes the non-electrostatic free energy penalty of 
adsorption per ion, $\alpha$, being linear in the fraction of occupied sites on a surface, as well as the change in 
the non-electrostatic interactions between the ions upon adsorption, formalized by the Flory-Huggins parameter $\chi$ 
and therefore quadratic in the fraction of occupied sites on a surface \cite{Har06, Mar16}. The last line 
in Eq.~(\ref{eq2}) is describing the mixing entropy of the adsorbed cations at neutral sites with probability 
$\eta\left(z\right)$. The values of $\alpha$ and $\chi$ are related to the specific chemistry of the 
two surfaces and the dissociation processes responsible for the charge regulation. In the case of charge-regulation 
by $\text{H}^{+}$ dissociation, $\alpha\approx\left(\text{pK}-\text{pH}\right)\ln 10$ can be tuned by 
changing the pH of the solution; pK corresponds to the equilibrium constant of the surface dissociation process \cite{Saf18}. 
Increasing $\alpha$, then promotes a favorable adsorption of protons onto the surface, while an increase in $\chi$ lowers 
the free energy of the system, so that an already adsorbed proton prefers the filling of a neighboring site.
The dimensionless parameters $\alpha$ and $\chi$ are phenomenological and their values are obtained from 
fitting the experimental data. Such an extension of the original charge regulation model by Ninham and 
Parsegian \cite{Nin71} was invoked in order to explain the details of an experimentally observed lamellar-lamellar phase 
transition in charged surfactant systems \cite{Dub98, Har06}. With both surfaces assumed to be chemically identical, and 
described by the same set $(\alpha, \chi)$ of phenomenological parameters, the equilibrium values for 
$\sigma^*\left(z\right)$ at the two surfaces are then determined by minimizing 
$\beta\widetilde\Omega\left[\sigma^*\right]$ in Eq.~(\ref{eq2}) with respect to $\sigma^*$.

\begin{figure*}[!t]
 \centering
 \includegraphics[width=15.4cm]{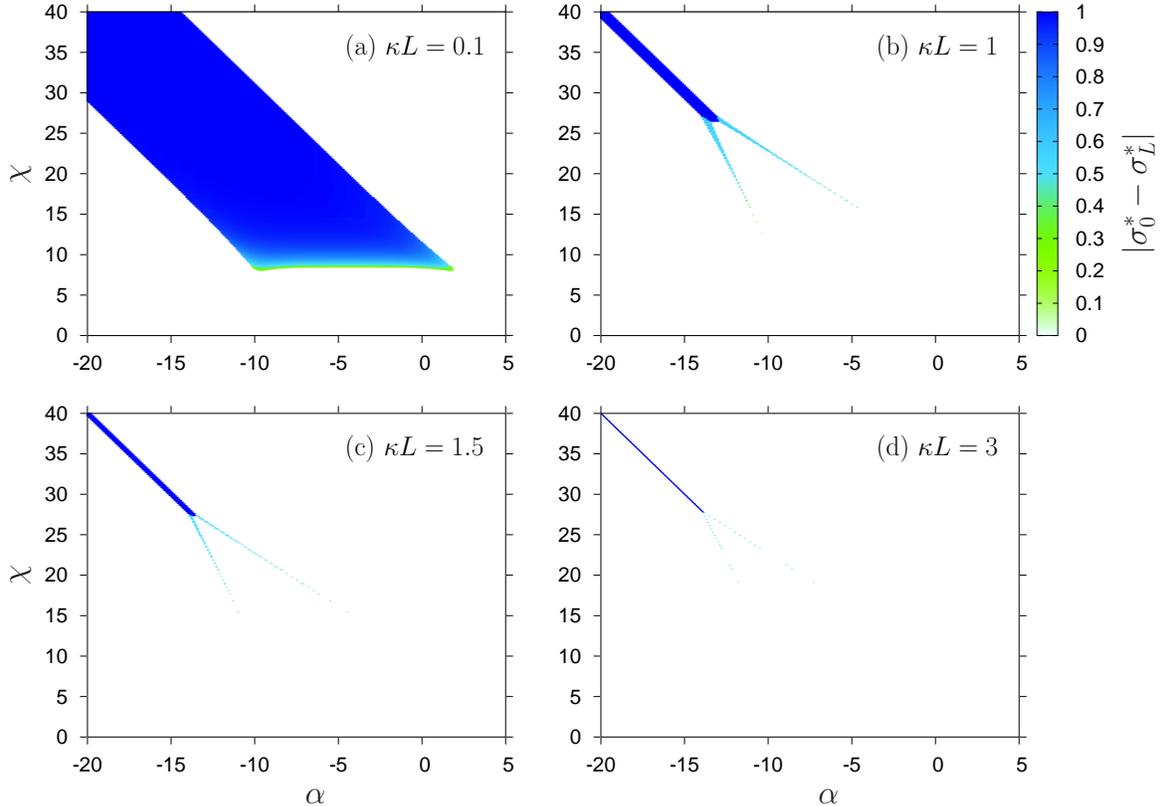}
 \caption{Absolute value of the difference of the dimensionless charge densities 
          $\sigma^*=\frac{a^2\sigma}{e}\in\left[-\frac{1}{2},\frac{1}{2}\right]$ at the two surfaces, 
          $\left|\sigma^*_0-\sigma^*_L\right|$, as function of $\alpha$ and
          $\chi$ (see Eq.~(\ref{eq2})) for (a) $\kappa L=0.1$, (b) $1$, (c) $1.5$, and (d) $3$. 
          The white regions in each figure correspond to a symmetric situation where $\sigma^*_0
          =\sigma^*_L$ whereas in the colored regions the equilibrium values of $\sigma^*_0$  and 
          $\sigma^*_L$ differ from one another.}
 \label{Fig2}
\end{figure*}

\section{Results and discussion} 
As mentioned earlier, the surface charge density profile $\sigma^{*}$ is laterally 
uniform on each surface, i.e., it may depend at most on 
$z=0,L$. In the following we 
use the notation $\sigma^{*}(0)=\sigma^{*}_0$ and $\sigma^{*}(L)=\sigma^{*}_L$. 
Both $\eta(0)$ and $\eta(L)$ can vary in the interval $[0,1]$ and for $\Theta=2$, which is the case
considered here, this corresponds to $\sigma^{*}\in\left[-\frac{1}{2},\frac{1}{2}\right]$ on each surface.
Figure~\ref{Fig2} shows the variation of the quantity $\left|\sigma^*_0-\sigma^*_L\right|$ with $\alpha$
and $\chi$ for gradually increasing separation $\kappa L$ between the surfaces. 
The parameters are varied in the intervals $\alpha\in[-20,5]$ and $\chi\in[0,40]$, which can be
considered as within the experimentally relevant regime \cite{Har06}.
Moreover realistic values $T=300\,\mathrm{K}$, $\varepsilon_r=80$ (water), and $a=1\,\mathrm{nm}$ are
used. Note that under these conditions $\kappa^{-1}\approx10\,\mathrm{nm}$ for an ionic strength 
$I=1\,\mathrm{mM}$. However, ionic strengths down to $\approx0.01\,\mathrm{mM}$ are used for experimental
studies in the present context \cite{Isr11}.

First, we consider the case $\kappa L=1$. As shown by Fig.~\ref{Fig2}(b), $\sigma^*_0$ and $\sigma^*_L$
are the same over a broad region (indicated by white) but \textit{not} everywhere. In the dark blue region, 
the charge asymmetry is the highest and close to unity, implying that the two surfaces are oppositely charged.
The line $\chi=-2\alpha$ passes through the middle of this region. 
Along this line, the solution is $\sigma^*_0=\sigma^*_L=0$ for $\alpha>\alpha_0$ ($\approx-13.2$ in
this case) and for $\alpha\lesssim\alpha_0$ the dark blue region appears. 
For $\alpha\gtrsim\alpha_0$ there are two more regions (one below the line $\chi=-2\alpha$ and one 
above) where the charge asymmetry is present albeit with a lower contrast 
$\left|\sigma^*_0-\sigma^*_L\right|$. 
These two tails (light blue or greenish) are not inter-connected but with decreasing $\alpha$ they 
thicken, come close to each other, and finally merge with the dark blue region. 
The charge contrast in each of these two regions increases with decreasing $\alpha$.
Below the dark blue region and the lower tail, the equilibrium states are symmetric with 
$\sigma^*_0=\sigma^*_L\gtrsim-\frac{1}{2}$ whereas above the dark blue region and the upper tail,  
they are symmetric with $\sigma^*_0=\sigma^*_L\lesssim\frac{1}{2}$. 
In between the two tails the states are $\sigma^*_0=\sigma^*_L\approx0$. 
In other words, $(\sigma^*_0,\sigma^*_L)$ changes from 
$(\gtrsim-\frac{1}{2},\gtrsim-\frac{1}{2})$ to $(\lesssim\frac{1}{2},\lesssim\frac{1}{2})$ 
across the dark blue region, from $(\gtrsim-\frac{1}{2},\gtrsim-\frac{1}{2})$ to $(\approx0,\approx0)$ 
across the lower tail, and from $(\approx0,\approx0)$ to $(\lesssim\frac{1}{2},\lesssim\frac{1}{2})$
across the upper tail.

\begin{figure*}[!t]
\centering{\includegraphics[width=16.0cm]{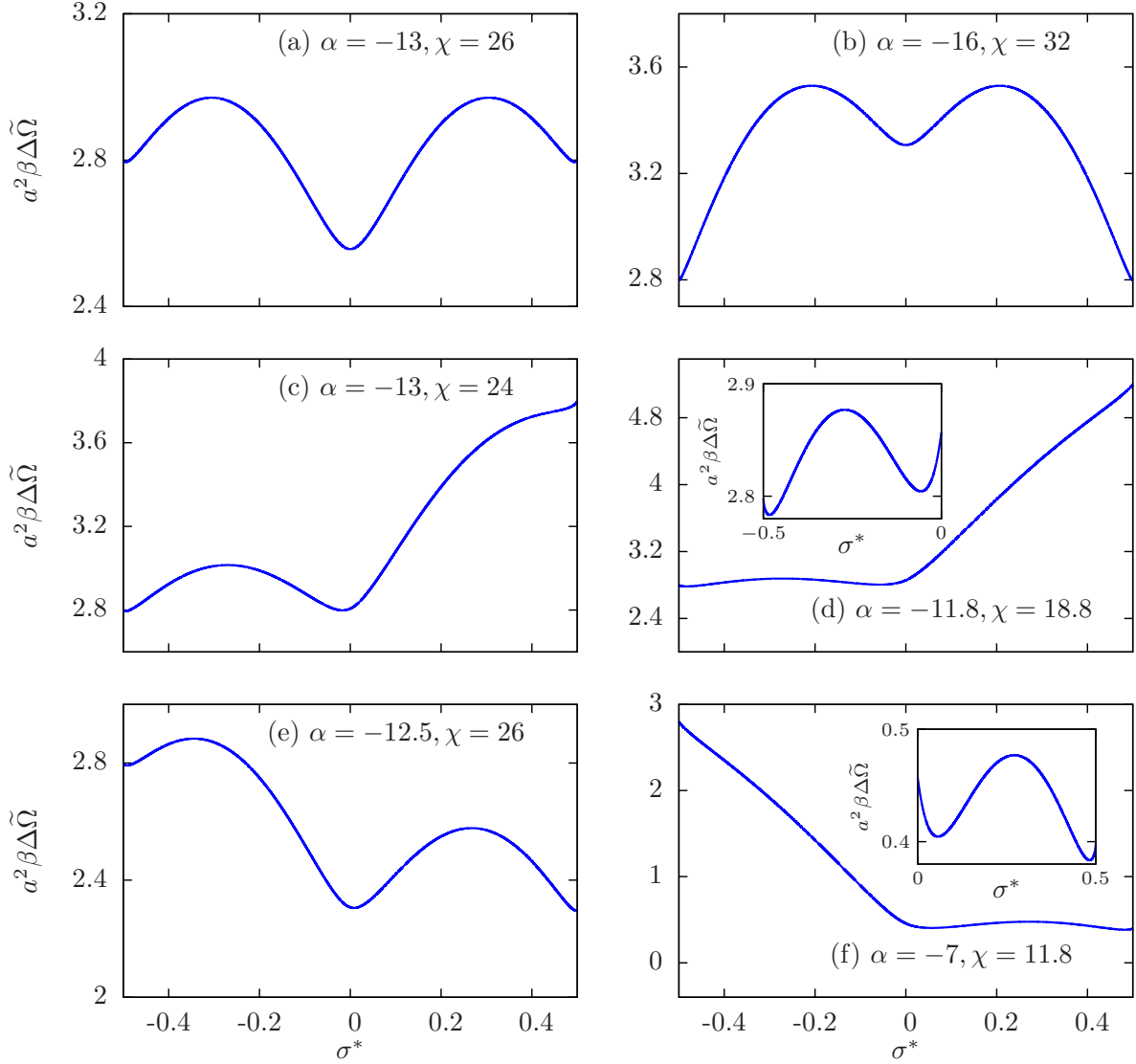}}
\caption{Variation of $\Delta\widetilde\Omega(\sigma^*)$ (see Eq.~(\ref{eq16a}) in the Appendix), which is obtained after 
         subtracting the bulk contribution from the grand potential functional $\widetilde\Omega$ per unit surface area 
         for a system consisting of a single charge-regulated surface in contact with an electrolyte and expressed in 
         the units of $1/\left(a^2\beta\right)$, as function of $\sigma^*$ for different combinations of the parameters 
         $\alpha$ and $\chi$. In all cases $\lambda=\beta e^2/\left(4\varepsilon\kappa a^2\right)\approx21.3$ is used.}
\label{Fig3}
\end{figure*}

With decreasing separation $L$ between the two surfaces, the dark blue region in Fig.~\ref{Fig2}
broadens and starts to dominate over the tails, making them hardly visible (see Fig.~\ref{Fig2}(a)).
With increasing separation all the regions shrink (see Figs.~\ref{Fig2}(c) and \ref{Fig2}(d)) and
the tails become increasingly difficult to be resolved numerically. 
In Fig.~\ref{Fig2} both the interaction parameters $\alpha$ and $\chi$ are sampled with a 
tenth of the thermal energy $k_BT=1/\beta$ because ion adsorption is governed by a 
competition with the bulk solvation free energy, which can usually be measured within a similar accuracy 
\cite{Mar83, Ine94}. However, the dark blue region seems to be very stable and it remains present even at 
$\kappa L=10$. Upon increasing $\kappa L$ from $0.1$ to $3$ both $\alpha_0$ and the width of the dark blue 
region decrease relatively fast, whereas for $\kappa L$ between $3$ and $10$, they hardly change.

\begin{figure}[!t]
\centering{\includegraphics[width=7cm]{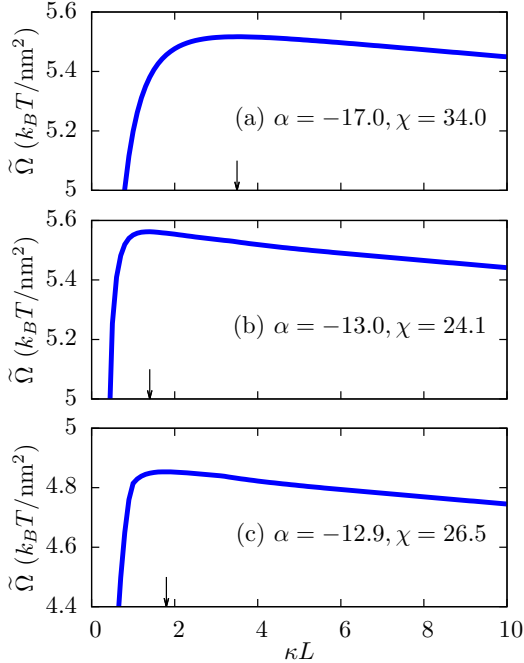}}
\caption{Variation of the effective interaction potential $\widetilde\Omega$ per unit surface area
         between two charge-regulated surfaces in the units of $k_BT/\mathrm{nm}^2$ as function of 
         the scaled separation $\kappa L$ for three different state points in Fig.~\ref{Fig2}(b) from 
         (a) the dark blue region, (b) the lower tail and (c) the upper tail. As shown by the plots, 
         the interaction energy increases initially, shows a maximum, and ultimately decreases monotonically. 
         The initial increase in the interaction energy corresponds to a negative effective force 
         at short distances $L$, implying that the interaction is attractive there. The value of 
         $\kappa L$ at the maximum of $\widetilde\Omega$ is indicated by arrows. As expected, the attraction 
         in the dark blue region of Fig.~\ref{Fig2} is stronger than in the other colored regions due to a 
         larger contrast in the surface charge densities.}
\label{Fig4}
\end{figure}

In order to explain these findings we consider a \textit{single} surface in contact with an 
electrolyte. This problem is analytically solvable and the solution shows that on the 
line $\chi=-2\alpha$, there are two equally deep (local) minima of the grand potential $\Omega(\sigma^*)$ at $\sigma^*_1$ and 
$\sigma^*_2=-\sigma^*_1$ (see the Appendix). For $\alpha\gtrsim-14$, $\sigma^*=0$ corresponds to the single global minimum of the grand 
potential; see Fig.~\ref{Fig3}(a). However, for $\alpha\lesssim-14$, the global minimum shifts to states with 
$\sigma^*_1\gtrsim-\frac{1}{2}$ and $\sigma^*_2=-\sigma^*_1\lesssim\frac{1}{2}$; see Fig.~\ref{Fig3}(b). 
For $\alpha\lesssim-14$, in the presence of a second surface, one surface acquires the 
charge density $\sigma^*=\sigma^*_1$ and the other $\sigma^*=\sigma^*_2=-\sigma^*_1$ because
the electrostatic attraction of two oppositely charged surfaces leads to a decrease of
the grand potential of the system. Similarly, the one-surface problem shows equally deep global minima 
of $\Omega(\sigma^*)$ at $\sigma^*\gtrsim-\frac{1}{2}$ and $\sigma^*\lesssim0$ for points in the upper 
part of the lower tail of Fig.~\ref{Fig2}(b) (e.g., see Fig.~\ref{Fig3}(c)). 
A second surface leads to charge densities of $\sigma^*\gtrsim-\frac{1}{2}$ on one surface and
of $\sigma^*\gtrsim0$ on the other such that the free energy cost in going upward the curve in
Fig.~\ref{Fig3}(c) is balanced by a reduction of the free energy due to electrostatic attraction.
As we go down the lower tail of Fig.~\ref{Fig2}(b), the one-surface problem shows two unequally deep 
minima at $\sigma^*\gtrsim-\frac{1}{2}$ and $\sigma^*\lesssim0$ for these points; see Fig.~\ref{Fig3}(d). 
Although for a single surface the minimum at $\sigma^*\gtrsim-\frac{1}{2}$ is slightly deeper than the
one at $\sigma^*\lesssim0$, the combination $\sigma^*_0=\sigma^*_L\gtrsim-\frac{1}{2}$ for two surfaces
would be too expensive due to strong electrostatic repulsion and $\sigma^*_0=\sigma^*_L\lesssim0$ is
also a state of higher free energy. The balance for two surfaces is obtained for the combination 
$(\gtrsim-\frac{1}{2},\lesssim0)$ by avoiding the repulsive interaction energy. 
As is shown in Figs.~\ref{Fig3}(e) and (f), a similar phenomenon occurs for the upper tail
in Fig.~\ref{Fig2}(b) except for the fact that there the equilibrium states are at 
$\sigma^*\gtrsim0$ and $\lesssim\frac{1}{2}$. With increasing separation $\kappa L$, the regions 
with charge asymmetry shrink because of a weaker electrostatic interaction due to screening.

\begin{figure}[!t]
\centering{\includegraphics[width=8.5cm]{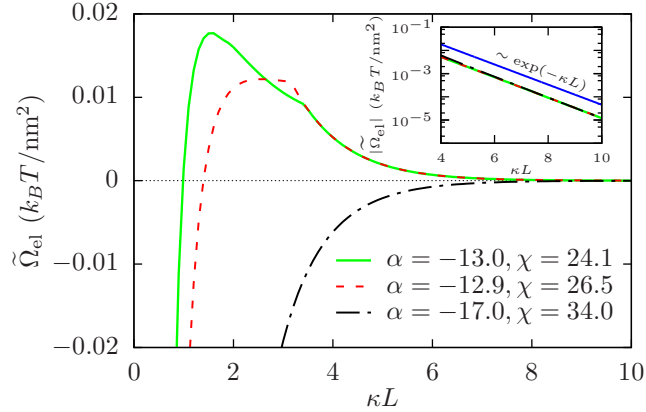}}
\caption{Electrostatic part $\widetilde\Omega_{\text{el}}$ of the effective interaction potential 
         $\widetilde\Omega$ per unit surface area between two charge-regulated surfaces in units of $k_BT/\mathrm{nm}^2$ as function of 
         the scaled separation $\kappa L$ for three different state points as in Fig.~\ref{Fig4}. For $\alpha=-17.0$ and $\chi=34.0$, 
         the interaction is attractive everywhere implied by the opposite signs of the charge densities at the two surfaces. 
         For the other two cases, the electrostatic interaction energies increase initially, show a maximum, 
         and then decay to zero. Both the curves show kinks within the repulsive part of the interaction which 
         are related to the discontinuities of the surface charge densities as functions of the wall separation 
         $\kappa L$ (see Fig.~\ref{Fig2}). As expected, the interaction decays exponentially to zero which is 
         confirmed by the semi-logarithmic plot in the inset.}
\label{Fig5}
\end{figure}

Once $\sigma^*(\mathbf{r})$ is known, the grand potential per unit surface area of the 
system can be obtained by evaluating $\widetilde\Omega\left[\sigma^*\right]$; see Eq.~(\ref{eq2}). 
The dependence of $\widetilde\Omega$ as function of the separation $\kappa L$ describes an effective 
interaction potential between the surfaces and is shown in Fig.~\ref{Fig4} for different combinations of the 
parameters $\alpha$ and $\chi$. In each case, $\widetilde\Omega$ increases initially with increasing $\kappa L$ 
and shows a maximum at some finite separation, typically well above the molecular length scale. Upon increasing $\kappa L$ further, 
the electrostatic interaction vanishes exponentially $\sim\exp(-\kappa L)$ (see Fig.~\ref{Fig5}) and the osmotic (or entropic) 
contribution ($=-2IL/\beta\sim L$) of the ions to $\widetilde\Omega$ dominates. The effective force per unit surface area
$-\partial\widetilde\Omega/\partial L$ is negative up to the distance $L$ of the maximum in Fig.~\ref{Fig4} 
and therefore, the interaction is attractive. This implies that the electrostatic attraction is sufficiently 
strong to overcome the repulsive osmotic pressure. At larger separations, however, the electrostatic 
interaction weakens and in the limit $L\rightarrow\infty$, the effective force per unit 
surface area equals the constant osmotic pressure $2I/\beta$. The occurrence of the maximum in Fig.~\ref{Fig4}(a) 
at a larger separation compared to the cases in Fig.~\ref{Fig4}(b) and (c) is related to enhanced electrostatic 
interactions due to the stronger asymmetry in the surface charge densities. Note that an effective interaction 
potential is a mesoscopic concept, which incorporates the energy and entropy balance of all microscopic degrees 
of freedom, e.g., the surface charge densities and the ion number density profiles, by minimization of the 
microscopic grand potential under the constraint of fixed mesoscopic degrees of freedom, e.g., the wall separation. 
In an even more microscopic (atomistic) approach, one could attempt to replace the parameters $\alpha$ and $\chi$ in 
favor of free energy contributions of the corresponding processes. In that sense an effective interaction energy 
always contains both energetic and entropic contributions.

The electrostatic part $\widetilde\Omega_\text{el}$ of the total interaction energy $\widetilde\Omega$ after 
subtracting the ideal osmotic (or entropic) contribution ($=-2IL/\beta\sim L$) of the ions and the surface tensions acting 
at the two solid-liquid interfaces, is shown as a function of the separation $\kappa L$ in Fig.~\ref{Fig5} for the same values
of the $\alpha$ and the $\chi$ parameters as in Fig.~\ref{Fig4}. For $\alpha=-17.0$ and $\chi=34.0$, which is 
a point on the line $\chi=-2\alpha$ in Fig.~\ref{Fig2}, the interaction is attractive due to the opposite charge densities at the 
two surfaces everywhere within the range of $\kappa L$ shown here. For the other two cases, the interactions are attractive 
at short distances $\kappa L$ and they become repulsive with increasing separation. These two curves show kinks corresponding to 
the discontinuities of the surface charge densities as functions of the wall separation $\kappa L$ (see Fig.~\ref{Fig2}). 
For example, if one considers the curve corresponding to $\alpha=-13.0$ and $\chi=24.1$, 
the two surfaces are oppositely charged up to $\kappa L\approx1.5$, then, from $\kappa L\approx1.5$ to $\kappa L\approx1.9$, the 
two surfaces carry equal charge densities ($\sigma^*_0=\sigma^*_L\lesssim\frac{1}{2}$), from $\kappa L\approx1.9$ to 
$\kappa L\approx3.4$ the surfaces adopt charge densities which are different in magnitude but have equal signs 
($\sigma^*_0\approx0$, $\sigma^*_L\lesssim\frac{1}{2}$), and finally, beyond $\kappa L\approx3.4$, both surfaces become equally 
charged ($\sigma^*_0=\sigma^*_L\approx0$). A similar phenomenon occurs for $\alpha=-12.9$ and $\chi=26.5$ where the interaction 
changes from attractive to repulsive at $\kappa L\approx2.9$, and at $\kappa L\approx3.2$ the charge densities at the 
two surfaces become equal. As expected, at large separations, the electrostatic interaction in all cases decay exponentially 
$\sim\exp(-\kappa L)$; see the inset of Fig.~\ref{Fig5}. Please note that the equilibrium states are characterized by a 
minimum of the total interaction potential $\widetilde\Omega$ as function of the wall separation $L$.
As $\widetilde\Omega(L)$ corresponds to the minimum of the grand potential functional $\Omega\left[\sigma^*,n_{\pm}\right]$
under the constraint of a fixed wall separation $L$, it is necessarily continuous with respect to $L$, but its derivatives 
with respect to $L$ may be discontinuous at first-order phase transitions. In the present work no first-order \emph{bulk} phase 
transitions are considered but the observed spontaneous symmetry breaking of the surface charge densities corresponds to first-order 
\emph{surface} phase transitions. Hence, kinks can occur only in the surface contribution $\widetilde\Omega_{\text{el}}(L)$, 
and they are hardly visible in the total interaction $\widetilde\Omega(L)$ (note the widely different scales in Figs.~\ref{Fig4} 
and \ref{Fig5}).

The inter-surface force is usually measured by using a surface forces apparatus (SFA), 
atomic force microscopy (AFM), or optical tweezers \cite{Isr11, Tre17}. In order to observe the anomalous 
attraction discussed in the preceding paragraph for a surface with an appropriate charge regulation behavior, 
one can either fix the distance between the surfaces and change the ionic strength of the solution or fix the 
ionic strength and vary the separation between the surfaces. For relatively small separations compared to 
the Debye length, $L \lesssim 1/\kappa$, the system is expected to exhibit a broad parameter range of surface 
charge asymmetry (see, e.g., the colored region in Fig.~\ref{Fig2}(a)). Charge asymmetry is also present for 
larger separations $L$, but the corresponding parameter range is smaller and more difficult to find (see Fig.~\ref{Fig2}). 
In order to avoid possible additional effects occurring at short separations $L$ in a real experimental setup, 
it is advisable to use low ionic strengths, i.e., large Debye lengths $1/\kappa$. For example, $I=0.1\,\mathrm{mM}$ 
in water leads to $1/\kappa\approx30\,\mathrm{nm}$, so that $\kappa L\approx0.1$ for a separation length 
$L\approx3\,\mathrm{nm}$, which is much larger than molecular dimensions and, therefore, a mean-field-like
theory as the one presented here is expected to work well \cite{Tre17}. Moreover, it is not necessary to go to 
such small values of $\kappa L$: between $\kappa L=0.1$ and $1$ (e.g., $\kappa L=0.5$), one can expect to have a 
sufficiently broad parameter range of surface charge asymmetry (see the colored regions in Fig.~\ref{Fig2}).
Possible candidates for the type of surfaces described here are biomolecules like lipids or proteins \cite{Isr11} or solid
colloidal particles (e.g., made of silica) grafted with particular surface groups (e.g., $-\text{NH}_{2}$ or $-\text{COOH}$) 
(see Refs.~\cite{Sol09, Yua17}). The parameter $\chi$ can be adjusted by means of an appropriate arrangement and 
density of surface groups. On the other hand, the parameter $\alpha$ can be tuned by changing counterion concentration, 
e.g., the pH, in the solvent. As mentioned earlier, the parameters $\alpha$ and $\chi$ can be obtained, e.g., 
by fitting experimentally measured profiles of the effective force. For example, for the synthetic cationic double-chain 
surfactant didodecyldimethylammonium ($\text{DDA}^+$) the values $\alpha=-7.4, \chi=14.75$ for bromide ($\text{Br}^-$) 
and $\alpha=-3.4,\chi=14.75$ for chloride ($\text{Cl}^-$) counterions have been obtained in Ref.~\cite{Har06}.
This demonstrates that the parameter ranges for $\alpha$ and $\chi$, for which surface charge asymmetry is predicted here,
are experimentally accessible, in particular in the case of low ionic strengths.

We finish our discussion by briefly commenting on the importance of the possible electrostatic 
attraction due to surface charge asymmetry in comparison with the van der Waals (vdW) attraction present in the system.
The vdW interaction is usually estimated in terms of the Hamaker coefficient \cite{Ham37}. 
For a pair of parallel planar silica surfaces interacting across water, the Hamaker coefficient is 
$A\approx4.8\times10^{-21}\,\mathrm{J}$ (see Ref.~\cite{Ber97}), so that the vdW attraction energy per unit 
surface area $-A/\left(12\pi L^2\right)\approx0.03\,k_BT/\mathrm{nm^2}$ for $L=1\,\mathrm{nm}$, which corresponds 
roughly to the thickness of the lines in Fig.~\ref{Fig4}. Hence the vdW interaction is qualitatively and quantitatively 
irrelevant for the effective interaction potentials considered in Fig.~\ref{Fig4}. The same can be expected for 
biological molecules, where the Hamaker coefficients are typically similar or smaller than $1\,k_BT$ (see
Ref.~\cite{Lec01}).

It is important to note that our findings are not restricted to the case $\Theta=2$.
For an asymmetric charge interval corresponding to $\Theta\neq2$, the colored regions of Fig.~\ref{Fig2}
shift in the $\alpha$-$\chi$ plane but the qualitative features remain the same.
In fact, we obtain asymmetric equilibrium states $\sigma^*_0\not=\sigma^*_L$ even for $\Theta=1$, 
where both surfaces can acquire only negative charges or remain uncharged; there the origin of
asymmetry is similar to the greenish regions in Fig.~\ref{Fig2}. 

\section{Conclusions}
In conclusion, our results clearly indicate that chemically identical charge-regulated surfaces in an
electrolyte \textit{are not} necessarily equally charged and need not repel each other.  
Even if the surfaces are equally charged at larger separations, their symmetry can become spontaneously broken with 
decreasing inter-surface distance and they can assume charge densities differing in magnitude as well as in sign. 
At short separations, but well-above the molecular scale, the resulting electrostatic 
attraction dominates over the repulsive osmotic (or entropic) pressure due to the ions and the 
vdW attraction between the surfaces. These findings contradict one of the fundamental assumptions commonly made in 
the application of the PB theory to chemically identical surfaces and puts it into an entirely new perspective. Since 
charge regulation is prevalent in most synthetic as well as natural colloids, including biomolecules, our findings are 
indeed expected to be relevant for a wide range of systems.


\begin{acknowledgments}
We thank Prof. P. A. Pincus for bringing up Ref.~\cite{Hen04} to our attention after the publication of 
this work, which we were not aware of before and have included in the reference list of this version.
\end{acknowledgments}


\appendix

\section{Single plate in contact with an electrolyte}

\subsection{Density functional}

Let us consider a single charge-regulated wall placed at $z=0$ in contact with an electrolyte 
solution of bulk ionic strength $I$ and spanning the space $z>0$. The charge density at the wall 
is denoted by $\sigma$ and the dimensionless charge density is given by
\begin{align}
\sigma^*=\frac{a^2\sigma}{e}=\eta-\frac{1}{\Theta}.
\label{eq0}
\end{align}
Please note that all variables used here and in the 
remainder have the same meaning as defined in the main text. After subtracting the bulk contribution
from the grand potential functional $\Omega$ corresponding to Eq.~(2) of the main text and afterwards
dividing by the surface area $\mathcal{A}$ of the wall one obtains
\begin{align}
 \frac{\beta\Delta\Omega\left(\sigma^*\right)}{\mathcal{A}}=& 
     -\int\limits_0^{\infty}dz\Bigg[2I\left(\cosh\left(\psi\left(z\right)\right)-1\right)
     +\frac{\varepsilon}{2\beta e^2}\left(\psi'\left(z\right)\right)^2\Bigg]
     +\frac{\sigma^*\psi\left(0\right)}{a^2}\notag\\
   & +\frac{1}{a^2}\left(-\alpha\eta
     -\frac{\chi}{2}\eta^2
     +\eta\ln\eta
     +\left(1-\eta\right)\ln\left(1-\eta\right)\right),
\label{eq1a}
\end{align}
where the dimensionless electrostatic potential $\psi$ satisfies the PB equation
\begin{align}
 \psi''\left(z\right)=\kappa^2\sinh\left(\psi\left(z\right)\right)
\label{eq2a}
\end{align}
subjected to the Dirichlet boundary condition $\psi(\infty)=0$ and to the Neumann boundary
condition
\begin{align}
 \psi'(0)=-\frac{\beta e\sigma}{\varepsilon}
         =-\frac{\beta e^2}{\varepsilon a^2}\left(\eta-\frac{1}{\Theta}\right)
         =-\frac{\beta e^2\sigma^*}{\varepsilon a^2}.
\label{eq3}
\end{align}
As usual $\psi'$ and $\psi''$ denote single and double derivatives with respect to $z$, respectively,
and $\eta=\sigma^*+\frac{1}{\Theta}$ according to Eq.~(\ref{eq0}).


\subsection{Grahame equation}

Multiplying both sides of Eq.~(\ref{eq2a}) by $\psi'$ one obtains
\begin{align*}
 \psi'\psi''=\kappa^2\sinh\left(\psi\right)\psi',
\end{align*}
which can be rewritten as
\begin{align}
 \frac{1}{2}\left(\left(\psi'\right)^2\right)'=\kappa^2\left(\cosh\left(\psi\right)\right)'.
\label{eq4}
\end{align}
Integrating Eq.~(\ref{eq4}) with respect to $z$ and using $\psi\left(\infty\right)=
\psi'\left(\infty\right)=0$ gives
\begin{align}
 \frac{1}{2}\left(\psi'\right)^2=\kappa^2\left(\cosh\left(\psi\right)-1\right),
\label{eq5}
\end{align}
which leads to
\begin{align}
 2I\left(\cosh\left(\psi\right)-1\right)=\frac{\varepsilon}{2\beta e^2}\left(\psi'\right)^2.
\label{eq5a}
\end{align}
For $z=0$, i.e., at the wall, Eq.~(\ref{eq5}) gives the Grahame equation \cite{Gra47}
\begin{align}
 \kappa^2\left(\cosh\left(\psi(0)\right)-1\right)=\frac{1}{2}\left(\psi'(0)\right)^2=\frac{\beta^2e^2\sigma^2}{2\varepsilon^2},
\label{eq6}
\end{align}
and therefore,
\begin{align}
 \psi(0)=\mathrm{sign}\left(\sigma\right)\operatorname{arcosh}\left(1+\frac{\beta^2e^2\sigma^2}{2\varepsilon^2\kappa^2}\right).
\label{eq7}
\end{align}
The sign of $\sigma$ and $\sigma^*$ are the same according to Eq.~(\ref{eq0}) and for brevity we define the
dimensionless parameter $\lambda=\frac{\beta e^2}{4\varepsilon\kappa a^2}$. With these,
Eq.~(\ref{eq7}) can be rewritten as
\begin{align}
 \psi(0)=\mathrm{sign}\left(\sigma^*\right)\operatorname{arcosh}\left(1+8\lambda^2\left(\sigma^*\right)^2\right).
\label{eq8}
\end{align}


\subsection{Electrostatic potential}
The PB equation for our setup is analytically solvable and its solution is well know \cite{Hun89, Rus89}:
\begin{align}
 \psi(z)=4\operatorname{artanh}\left(\gamma\exp\left(-\kappa z\right)\right);~~~~~\gamma=\tanh\left(\frac{\psi(0)}{4}\right).
\label{eq9}
\end{align}
Taking the derivative with respect to $z$, one obtains
\begin{align}
 \psi'(z)=-4\kappa\gamma\frac{\exp\left(-\kappa z\right)}{1-\gamma^2\exp\left(-2\kappa z\right)}.
\label{eq10}
\end{align}
Therefore,
\begin{align}
 \int\limits_0^{\infty}dz\left(\psi'(z)\right)^2&=16\kappa^2\gamma^2\int\limits_0^{\infty}dz
         \frac{\exp\left(-2\kappa z\right)}{\left(1-\gamma^2\exp\left(-2\kappa z\right)\right)^2}\notag\\
        &=8\kappa\int\limits_0^{\infty}dz
         \frac{2\kappa\gamma^2\exp\left(-2\kappa z\right)}{\left(1-\gamma^2\exp\left(-2\kappa z\right)\right)^2}\notag\\
        &=8\kappa\int\limits_0^{\infty}dz
         \left(\frac{d}{dz}\frac{-1}{1-\gamma^2\exp\left(-2\kappa z\right)}\right)\notag\\
        &=8\kappa\left|\frac{-1}{1-\gamma^2\exp\left(-2\kappa z\right)}\right|_{z=0}^{\infty}\notag\\
        &=8\kappa\left(-1+\frac{1}{1-\gamma^2}\right)\notag\\
        &=\frac{8\kappa\gamma^2}{1-\gamma^2}.
\label{eq11}
\end{align}
The parameter $\gamma$ is determined by using the boundary condition relating the electric displacement
vector to the charge density at the wall. Combining Eqs.~(\ref{eq3}) and (\ref{eq10}), one obtains
\begin{align}
 \psi'(0)=-4\kappa\frac{\gamma}{1-\gamma^2}=-\frac{\beta e\sigma}{\varepsilon},
\label{eq12}
\end{align}
which leads to
\begin{align}
 \frac{\gamma}{1-\gamma^2}=\frac{\beta e^2\sigma^*}{4\kappa\varepsilon a^2}=\lambda\sigma^*.
\label{eq13}
\end{align}
Solving Eq.~(\ref{eq13}) for $\gamma$ and inserting it in Eq.~(\ref{eq11}), one finally arrives at
\begin{align}
 \int\limits_0^{\infty}dz\left(\psi'(z)\right)^2=4\kappa\left(-1+\sqrt{1+4\lambda^2\left(\sigma^*\right)^2}\right).
\label{eq14}
\end{align}


\subsection{Grand potential}

Using Eqs.~(\ref{eq5a}) and (\ref{eq8}) in Eq.~(\ref{eq1a}), one can write
\begin{align}
 \frac{\beta\Delta\Omega\left(\sigma^*\right)}{\mathcal{A}}=& 
     -\int\limits_0^{\infty}dz\frac{\varepsilon}{\beta e^2}\left(\psi'\right)^2
     +\frac{\sigma^*}{a^2}\mathrm{sign}\left(\sigma^*\right)
     \operatorname{arcosh}\left(1+8\lambda^2\left(\sigma^*\right)^2\right)\notag\\
   & +\frac{1}{a^2}\left(-\alpha\eta
     -\frac{\chi}{2}\eta^2
     +\eta\ln\eta
     +\left(1-\eta\right)\ln\left(1-\eta\right)\right).
\label{eq15}
\end{align}
Further, using Eq.~(\ref{eq14}), Eq.~(\ref{eq15}) can be rewritten as
\begin{align}
 \frac{\beta\Delta\Omega\left(\sigma^*\right)}{\mathcal{A}}=\frac{1}{a^2} 
     \Bigg[&\frac{1}{\lambda}\left(1-\sqrt{1+4\lambda^2\left(\sigma^*\right)^2}\right)
     +\left|\sigma^*\right|\operatorname{arcosh}\left(1+8\lambda^2\left(\sigma^*\right)^2\right)\notag\\
   & -\alpha\eta
     -\frac{\chi}{2}\eta^2
     +\eta\ln\eta
     +\left(1-\eta\right)\ln\left(1-\eta\right)\Bigg],
\label{eq16}
\end{align}
where the relation $\frac{4\kappa\varepsilon a^2}{\beta e^2}=\frac{1}{\lambda}$ is used.


\subsection{Symmetric charge interval ($\Theta=2$)}

As mentioned in the main text, $\Theta=2$ corresponds to a symmetric charge interval. For this case,
$\eta=\sigma^*+\frac{1}{2}$ according to Eq.~(\ref{eq0}) and using this, Eq.~(\ref{eq16}) can be written as:
\begin{align}
 \beta\Delta\widetilde\Omega\left(\sigma^*\right)=\frac{1}{a^2} 
     \Bigg[&\frac{1}{\lambda}\left(1-\sqrt{1+4\lambda^2\left(\sigma^*\right)^2}\right)
     +\left|\sigma^*\right|\operatorname{arcosh}\left(1+8\lambda^2\left(\sigma^*\right)^2\right)\notag\\
   & -\alpha\left(\frac{1}{2}+\sigma^*\right)
     -\frac{\chi}{2}\left(\frac{1}{2}+\sigma^*\right)^2\notag\\
   & +\left(\frac{1}{2}+\sigma^*\right)\ln\left(\frac{1}{2}+\sigma^*\right)
     +\left(\frac{1}{2}-\sigma^*\right)\ln\left(\frac{1}{2}-\sigma^*\right)\Bigg],
\label{eq16a}
\end{align}
where $\Delta\widetilde\Omega\left(\sigma^*\right)=\frac{\Delta\Omega\left(\sigma^*\right)}{\mathcal{A}}$ is the 
free energy per unit surface area. Clearly, $\beta\Delta\widetilde\Omega\left(\sigma^*\right)$ in Eq.~(\ref{eq16a}) 
is symmetric about $\sigma^*=0$, 
i.e., $\beta\Delta\widetilde\Omega\left(-\sigma^*\right)=\beta\Delta\widetilde\Omega\left(\sigma^*\right)$, provided the 
condition
\begin{align}
 \left(2\alpha+\chi\right)\sigma^*=0
\label{eq17}
\end{align}
is fulfilled. 
According to this condition, on the line $\chi=-2\alpha$ two states with $\sigma^*_1$ and 
$\sigma^*_2=-\sigma^*_1$ correspond to the same value of $\beta\Delta\widetilde\Omega(\sigma^*)$. 
Therefore, if a state with $\sigma^*_1$ corresponds to the global minimum of $\beta\Delta\widetilde\Omega(\sigma^*)$,
there will be another state with $\sigma^*_2=-\sigma^*_1$ with the same minimum, i.e., the two states with
$\sigma^*_1$ and $\sigma^*_2$ coexist. 
As shown in Fig.~3 of the main text, for $\alpha\gtrsim-14$ and $\chi=-2\alpha$ the global minimum corresponds to
$\sigma^*=-\sigma^*=0$ (see Fig.~3(a)) whereas for $\alpha\lesssim-14$, it shifts to 
$\sigma^*_1\gtrsim-\frac{1}{2}$ and $\sigma^*_2=-\sigma^*_1\lesssim\frac{1}{2}$ (see Fig.~3(b)).



\end{document}